# Estimation of Operational Risk Capital Charge under Parameter Uncertainty

**Pavel V. Shevchenko**
Principal Research Scientist, CSIRO Mathematical and Information Sciences, Sydney,
Locked Bag 17, North Ryde, NSW, 1670, Australia.
e-mail: Pavel.Shevchenko@csiro.au

**Abstract**
Many banks adopt the Loss Distribution Approach to quantify the operational risk capital charge under Basel II requirements. It is common practice to estimate the capital charge using the 0.999 quantile of the annual loss distribution, calculated using point estimators of the frequency and severity distribution parameters. The uncertainty of the parameter estimates is typically ignored. One of the unpleasant consequences for the banks accounting for parameter uncertainty is an increase in the capital requirement. This paper demonstrates how the parameter uncertainty can be taken into account using a Bayesian framework that also allows for incorporation of expert opinions and external data into the estimation procedure.

# Introduction

Under the Basel II requirements BIS (2006) banks are required to quantify the capital charge for operational risk. Many banks have adopted the Loss Distribution Approach (LDA), where frequency and severity of operational losses for each risk cell (in the matrix of eight business lines times seven event types) are estimated over a one year period. Then the capital is estimated using the 0.999 quantile of the distribution for the total annual loss in the bank. Under the LDA, the annual loss in a risk cell is

$$Z = \sum_{i=1}^{N} X_i, \qquad (1)$$

where $N$ is the annual number of events, modelled as a random variable from some discrete distribution with a density $p(.|\boldsymbol{\alpha})$, and $X_i$, $i=1,...,N$, are the severities of the events modelled as independent random variables from a continuous distribution with a density $f(.|\boldsymbol{\beta})$. Here, $\boldsymbol{\alpha}$ and $\boldsymbol{\beta}$ are distribution parameters. The frequency $N$ and severities $X_i$ are assumed to be independent.

Estimation of the frequency and severity distributions is a challenging task but it will not be discussed here. There is an extensive literature on modelling the risk process (1), see e.g. McNeil, Frey and Embrechts (2005). In this paper, we assume that the distribution type is known and address the issue of the parameter uncertainty (parameter risk). The parameters $\boldsymbol{\alpha}$ and $\boldsymbol{\beta}$ are unknown and statistical estimation of the parameters would involve some uncertainty due to the finite sample size of the observed data. Typically, the capital is estimated using point estimators for $\boldsymbol{\alpha}$ and $\boldsymbol{\beta}$, e.g. Maximum Likelihood Estimators (MLEs). Although the parameter estimates are uncertain, in practice this uncertainty is commonly ignored in the estimation of the operational risk capital charge. Taking parameter uncertainty into account is typically regarded as very difficult and time consuming calculation. One unpleasant consequence for a bank accounting for parameter uncertainty is an increase in the capital charge estimate when compared to the capital based on the point estimators for model parameters.

In our opinion, this subject deserves more attention. Only a few papers mention this issue. Frachot, Moudoulaud and Roncalli (2004) discussed the accuracy of the capital charge by constructing a confidence interval for the 0.999 quantile of the annual loss distribution. Mignola and Ugoccioni (2006) assessed the standard deviation of the 0.999 quantile using a first order Taylor expansion approximation. They demonstrated that even small uncertainty in the parameters can lead to profound uncertainty in the 0.999 quantile. It is critical to account for parameter uncertainty, otherwise capital charge could be underestimated significantly. The official position of BIS (2006) regarding accounting for parameter uncertainty is not clear. However, in practice, we experienced that often local regulators have concerns about the correctness of the parameter point estimators used by a bank. Appropriate accounting for parameter uncertainty would help a bank to resolve these concerns. In this paper, we present a convenient and practical approach (not difficult to implement) to account for parameter uncertainty using Bayesian inference.

Bayesian inference is a statistical technique well suited to model parameter uncertainty. There is a broad literature covering Bayesian inference and its applications for the insurance industry as well as other areas; see e.g. Berger (1985). The method allows for expert opinions or external data to be incorporated into the analysis via specifying so-called prior distributions for model parameters. In the application to operational risk, the method is relatively new, see Shevchenko and Wüthrich (2006). Below we focus on the application of the method to the estimation of the operational risk capital, taking into account the parameter uncertainty. Accounting for parameter uncertainty is critical for operational risk due to the limited internal data for some risks, very high quantile level and long time horizon used for capital estimation.

## Bayesian inference

For simplicity, consider one risk cell only. Under model (1), frequencies $N$ are iid from $p(.|\boldsymbol{\alpha})$ and severities $X_i$ are iid from $f(.|\boldsymbol{\beta})$, where $\boldsymbol{\theta} = (\boldsymbol{\alpha}, \boldsymbol{\beta})$ is a vector of distribution parameters. Denote the density of the annual loss $Z$, conditional on parameters $\boldsymbol{\theta}$, as $g(Z|\boldsymbol{\theta})$. Given $\boldsymbol{\theta}$, this distribution is usually calculated numerically by Monte Carlo (MC), Panjer recursion or Fast Fourier Transform methods. Typically, given observations, the MLE $\hat{\boldsymbol{\theta}}$ is used as the "best fit" point estimators for $\boldsymbol{\theta}$. Then the annual loss distribution for the next year is estimated as $g(Z|\hat{\boldsymbol{\theta}})$ and its 0.999 quantile, $\hat{Q}_{0.999}$, is used for the capital charge calculation.

However, the parameters $\boldsymbol{\theta}$ are unknown and can be modeled as random variable with the density $\pi(\boldsymbol{\theta}|\mathbf{Y})$. Here, $\mathbf{Y}$ is a vector of all loss events (frequencies and severities) used in the estimation procedure. Then, the marginal (predictive) distribution of the annual loss $Z$, taking into account the parameter uncertainty, is

$$g(Z|\mathbf{Y}) = \int g(Z|\boldsymbol{\theta}) \times \pi(\boldsymbol{\theta}|\mathbf{Y}) d\boldsymbol{\theta}. \qquad (2)$$

The capital charge, accounting for the parameter uncertainty, should be based on the 0.999 quantile, $\hat{Q}^B_{0.999}$, of the predictive distribution $g(Z|\mathbf{Y})$. The above formula can be viewed as a weighted average of distributions $g(Z|\boldsymbol{\theta})$ with weights $\pi(\boldsymbol{\theta}|\mathbf{Y})$. If there is no uncertainty, then $\Pr(\boldsymbol{\theta} = \boldsymbol{\theta}_0) = 1$ and $g(Z|\mathbf{Y}) = g(Z|\boldsymbol{\theta}_0)$.

It is convenient to model parameter uncertainty using Bayesian inference. Consider a random vector of loss frequencies $N_m$ and severities $X_i$ over $M$ years: $\mathbf{Y} = (N_1, ..., N_M, X_1, ..., X_n)$, where $n = \sum_{m=1}^{M} N_m$. Denote the density of $\mathbf{Y}$, given parameters $\boldsymbol{\theta}$, as $h(\mathbf{Y}|\boldsymbol{\theta})$. Then according to Bayes' theorem

$$h(\mathbf{Y}, \boldsymbol{\theta}) = h(\mathbf{Y}|\boldsymbol{\theta})\pi(\boldsymbol{\theta}) = \pi(\boldsymbol{\theta}|\mathbf{Y})h(\mathbf{Y}), \qquad (3)$$

where
- $\pi(\boldsymbol{\theta})$ is the density of parameters, a so-called prior distribution. $\pi(\boldsymbol{\theta})$ depends on a set of further parameters that are called hyper-parameters, omitted hereafter for simplicity of notation;
- $\pi(\boldsymbol{\theta}|\mathbf{Y})$ is the density of parameters given data $\mathbf{Y}$, a so-called posterior distribution;
- $h(\mathbf{Y}, \boldsymbol{\theta})$ is the joint density of data and parameters;

- $h(\mathbf{Y}|\boldsymbol{\theta}) = \prod_{m=1}^{M} p(N_m|\boldsymbol{\alpha}) \prod_{i=1}^{n} f(X_i|\boldsymbol{\beta})$ is a likelihood function, density of $\mathbf{Y}$ given parameters $\boldsymbol{\theta}$ (here, we use the assumption that frequencies and severities are independent given $\boldsymbol{\theta}$);
- $h(\mathbf{Y}) = \int h(\mathbf{Y}|\boldsymbol{\theta}) \pi(\boldsymbol{\theta}) d\boldsymbol{\theta}$ is a marginal density of $\mathbf{Y}$.

Hereafter, we consider continuous $\pi(\boldsymbol{\theta})$. If $\pi(\boldsymbol{\theta})$ is a discrete distribution, then the relevant integrations should be replaced with summations. Given that frequencies and severities are independent, $\boldsymbol{\alpha}$ and $\boldsymbol{\beta}$ can be assumed independent and their distributions can be estimated marginally using observed frequencies and severities separately. Note that if data are reported above some level then $\boldsymbol{\alpha}$ and $\boldsymbol{\beta}$ are dependent and full joint likelihood should be used for estimation, see Luo, Shevchenko and Donnelly (2007). Using (3), the posterior distribution can be written as

$$\pi(\boldsymbol{\theta}|\mathbf{Y}) \propto h(\mathbf{Y}|\boldsymbol{\theta}) \pi(\boldsymbol{\theta}), \qquad (4)$$

where the normalization constant $h(\mathbf{Y})$ is omitted. Hereafter, $\propto$ is used to indicate that distributions are equivalent up to a normalization factor. Thus, the posterior distribution can be viewed as the product of a prior knowledge with a likelihood function of observed data. In practice, initially, we start with the prior distribution $\pi(\boldsymbol{\theta})$ identified by expert opinions or external data only. Then, the posterior distribution $\pi(\boldsymbol{\theta}|\mathbf{Y})$ is calculated, using (4), when actual data are observed. If there is a reason, for example a new control policy introduced in a bank, then this posterior distribution can be adjusted by an expert and treated as the prior distribution for subsequent observations (alternatively, one can try to adjust the observed data as if a policy was introduced before and repeat the estimation procedure). Often the posterior distributions are used to find point estimators of the parameters using the mean, median or mode of the posterior. However, for the purposes of accounting for parameter uncertainty using (2), the whole posterior distribution is required. In the case of no prior knowledge about parameters (or if the estimation inferences should be based on observed data only), one can use non-informative improper constant prior to rely on observations only, i.e. assume that

$$\pi(\boldsymbol{\theta}|\mathbf{Y}) \propto h(\mathbf{Y}|\boldsymbol{\theta}). \qquad (5)$$

Note that in this case, the mode of the posterior distribution is a point estimator of the parameters equivalent to the MLEs and inferences based on posterior distribution are equivalent to the inferences based on the maximum likelihood method. Alternatively, one can use e.g. constant priors defined on a very wide interval. It is important to note that, in the limit of a large number of observations, $\hat{Q}_{0.999}^B$ converges to $\hat{Q}_{0.999}$, because the variances of the posterior distributions approach zero (if the prior distribution is continuous at the "true" value of parameters).

Another approach under Bayesian framework to account for parameter uncertainty is to consider a quantile $Q_{0.999}(\boldsymbol{\theta})$ of the conditional annual loss distribution $g(Z|\boldsymbol{\theta})$. Then, given that $\boldsymbol{\theta}$ is distributed from $\pi(\boldsymbol{\theta}|\mathbf{Y})$, one can find the distribution for $Q_{0.999}(\boldsymbol{\theta})$ and form a predictive interval to contain the true value with some probability $q$. This is similar to forming a confidence interval in the frequentist approach using distribution of $\hat{Q}_{0.999}(\hat{\boldsymbol{\theta}})$, where $\hat{\boldsymbol{\theta}}$ is treated as random (usually, Gaussian distribution estimated by maximum likelihood method is assumed for $\hat{\boldsymbol{\theta}}$. Often, if derivatives can be calculated efficiently, the variance of $\hat{Q}_{0.999}(\hat{\boldsymbol{\theta}})$ is simply estimated via error propagation method and a first order Taylor expansion). Under this approach, one can argue that the conservative estimate of the capital charge accounting for parameter uncertainty should be based on the upper bound of the constructed interval. This approach is more computationally demanding if MC method is used. Also, specification of the confidence level $q$ is required - it might be difficult to argue that commonly used confidence level $q = 0.95$ (or less) is good enough for 0.999 quantile. In operational risk, it seems more appealing to estimate the whole predictive distribution (2) for the annual loss and use it for capital charge quantification.

## The capital charge

For the purposes of the regulatory capital calculations of operational risk, the annual loss distribution (in particular its 0.999 quantile as a risk measure) should be quantified for each risk cell in the bank. Consider a risk cell in the bank. Assume that the frequency $p(.|\boldsymbol{\alpha})$ and severity $f(.|\boldsymbol{\beta})$ densities for the cell are chosen. Also, suppose that the posterior distribution $\pi(\boldsymbol{\theta}|\mathbf{Y})$, $\boldsymbol{\theta} = (\boldsymbol{\alpha}, \boldsymbol{\beta})$, is estimated using (4), in the following sections some closed form solutions are presented. Then, under model (1), the predictive annual loss distribution (2) in the cell can be calculated using, for example, MC procedure with the following logical steps:

**Step1**. For a given risk simulate the risk parameters $\boldsymbol{\theta} = (\boldsymbol{\alpha}, \boldsymbol{\beta})$ from the posterior distribution $\pi(\boldsymbol{\theta}|\mathbf{Y})$.

**Step2**. Given $\boldsymbol{\alpha}$ and $\boldsymbol{\beta}$ from Step 1: simulate the annual number of events $N$ from $p(.|\boldsymbol{\alpha})$; simulate severities $X_n, n = 1,...,N$ from $f(.|\boldsymbol{\beta})$; calculate the annual loss $Z = \sum_{n=1}^{N} X_n$.

**Step3**. Repeat Steps 1-2 $K$ times to build a sample of possible annual losses $Z_1,...,Z_K$. The 0.999 quantile $\hat{Q}_{0.999}^B$ and other distribution characteristics are estimated using the above simulated sample in the usual way.

*Remarks:*
- Assume that the sample $Z_1,...,Z_K$ is sorted into ascending order $Z_1 \leq ... \leq Z_K$, then the quantile $Q_q$ is estimated by $Z_{\lfloor Kq+1 \rfloor}$. Here, $\lfloor \ \rfloor$ denotes rounding downward.
- Numerical error (due to the finite number of simulations $K$) in the quantile estimator can be assessed by forming a conservative confidence interval $[Z_r, Z_s]$ utilizing the fact that the number of samples not exceeding the quantile $Q_q$ has a Binomial distribution with parameters $q$ and $K$ (i.e. with mean $= Kq$ and var $= Kq(1-q)$). Approximating the Binomial by the Normal distribution leads to a simple formula for the conservative confidence interval:

$$r = \lfloor l \rfloor, \ l = Kq - z_{(1+\gamma)/2}\sqrt{Kq(1-q)},$$
$$s = \lceil u \rceil, \ u = Kq + z_{(1+\gamma)/2}\sqrt{Kq(1-q)}, \quad (6)$$

where $\gamma$ is a confidence level, $\lceil \ \rceil$ denotes rounding upwards and $z_{(1+\gamma)/2}$ is a $(1+\gamma)/2$ quantile of the standard Normal distribution. The above formula works very well for $Kq(1-q) \geq 50$ approximately and, in particular, for large quantiles with $q \leq 0.9995$ when the number of MC simulations $K \geq 10^5$. One-sided confidence intervals can be formed easily.

- A large number of simulations, typically $K \geq 10^5$, should be used to achieve a good numerical accuracy for 0.999 quantile. One of the approaches is to continue simulations until a desired numerical accuracy is achieved.

In the above procedure the risk profiles **α** and **β** are simulated from their posterior distribution for each simulation. Thus, we model both the process risk (process uncertainty), which comes from the fact that frequencies $N$ and severities $X_n$ are random variables, and the parameter risk (parameter uncertainty), which comes from the fact that we do not know the true values of **α**, **β**. The annual loss distribution calculated in the above is the predictive distribution (2). To calculate the conditional distribution $g(Z|\hat{\boldsymbol{\theta}})$ and its quantile $\hat{Q}_{0.999}$ using parameter point estimators $\hat{\boldsymbol{\theta}}$, Step1 should be simply replaced with setting $\boldsymbol{\theta} = \hat{\boldsymbol{\theta}}$ for all simulations $k = 1,...,K$. Thus, MC calculations of $\hat{Q}^B_{0.999}$ and $\hat{Q}_{0.999}$ are similar, given that $\pi(\boldsymbol{\theta}|\mathbf{Y})$ is known. If $\pi(\boldsymbol{\theta}|\mathbf{Y})$ is not known in closed form then it can be estimated efficiently using Gaussian approximation (described in the next section) or available MCMC algorithms, see Peters and Sisson (2006).

According to the Basel II requirements BIS (2006), the final bank capital should be calculated as the sum of the risk measures in the risk cells if the bank's model cannot account for correlations between risks accurately. If this is the case, then Steps 1-3 should be performed for each risk cell to estimate their risk measures separately. Then these are summed to estimate the total bank capital. Of course, adding quantiles over the risk cells to find the quantile of the total loss distribution can be too conservative, as it is equivalent to the assumption of perfect dependence between risks. An attractive way to model the dependence between different risks is via dependence between risk profiles $\boldsymbol{\theta}_j$, $j = 1,...,J$ (here, the subscript *j* is used to indicate risk *j*) using an appropriate dependence structure (copula), e.g. considering risk parameters evolving in time and dependent. Also, common shock processes can be used to model dependence between risk cells. For further information we refer to McNeil, Frey and Embrechts (2005). Accurate quantification of the dependencies between the risks is a difficult task and is an open field for future research.

## Posterior distribution

In some cases the posterior distribution (4) can be found analytically. This is the case for many so called conjugate prior distributions, where prior and posterior distributions have the same type. Some of these distributions are used in the numerical examples in this paper. The formulas for the corresponding posterior distributions will be presented in the following sections and more details can be found in Shevchenko and Wüthrich (2006). However, in general, the posterior distribution should be calculated numerically using Markov Chain Monte Carlo methods, see Peters and Sisson (2006).

Sometimes the following simple approximation can be used. Let $\hat{\boldsymbol{\theta}}$ be a mode of the posterior distribution $\pi(\boldsymbol{\theta}|\mathbf{Y})$. It can be used as a point estimator of the parameters and is equivalent to the MLEs in the case of constant priors. One can find a mode by maximizing $\pi(\boldsymbol{\theta}|\mathbf{Y})$ using standard optimization algorithms. Then, in the case of a continuous prior at $\hat{\boldsymbol{\theta}}$, the posterior distribution can be approximated using a second order Taylor series expansion around $\hat{\boldsymbol{\theta}}$ as

$$\ln \pi(\boldsymbol{\theta}|\mathbf{Y}) \approx \ln \pi(\hat{\boldsymbol{\theta}}|\mathbf{Y}) + \frac{1}{2}\sum_{i,j}\left.\frac{\partial^2 \ln \pi(\boldsymbol{\theta}|\mathbf{Y})}{\partial \theta_i \partial \theta_j}\right|_{\boldsymbol{\theta}=\hat{\boldsymbol{\theta}}}(\theta_i - \hat{\theta}_i)(\theta_j - \hat{\theta}_j). \qquad (7)$$

Here, we used the fact that the first order partial derivatives are zero at $\boldsymbol{\theta} = \hat{\boldsymbol{\theta}}$ because $\hat{\boldsymbol{\theta}}$ was chosen to be a mode of $\pi(\boldsymbol{\theta}|\mathbf{Y})$. Thus $\pi(\boldsymbol{\theta}|\mathbf{Y})$ is approximately a multivariate Normal distribution with mean $\hat{\boldsymbol{\theta}}$ and covariance matrix $\boldsymbol{\Sigma}$ calculated as the inverse of the Fisher information matrix $\mathbf{I}$ with the elements

$$(\mathbf{I})_{ij} = -\left.\partial^2 \ln \pi(\boldsymbol{\theta}|\mathbf{Y})/\partial \theta_i \partial \theta_j\right|_{\boldsymbol{\theta}=\hat{\boldsymbol{\theta}}}. \qquad (8)$$

In the case of improper constant priors, i.e. formula (5), this approximation is equivalent to the maximum likelihood estimation of the parameters and their covariances.

In practice, it is not unusual to restrict parameters. In this case the posterior distribution will be a truncated version of the posterior distribution in the unrestricted case. For example, if from expert opinions or external data we identified that $\boldsymbol{\theta}$ is restricted to some range $[\boldsymbol{\theta}_L, \boldsymbol{\theta}_H]$ then the posterior distribution will have the same type as in the unrestricted case but truncated outside this range.

## Poisson distribution (frequency)

Suppose that, given $\lambda$, the annual counts $\mathbf{N} = (N_1, ..., N_n)$ are independent random variables from Poisson distribution, $Poisson(\lambda)$, with a density

$$f(N|\lambda) = e^{-\lambda}\frac{\lambda^N}{N!}, \quad \lambda > 0. \qquad (9)$$

Then the likelihood function is $h(\mathbf{N}|\lambda) = \prod_{i=1}^{n} e^{-\lambda}\lambda^{N_i}/N_i!$. If the prior distribution of $\lambda$ is $Gamma(\alpha, \beta)$:

$$\pi(\lambda) = \frac{(\lambda/\beta)^{\alpha-1}}{\Gamma(\alpha)\beta} \exp(-\lambda/\beta), \quad \alpha > 0, \quad \beta > 0, \tag{10}$$

with mean $= \alpha\beta$ and var $= \alpha\beta^2$, then using (4), the posterior distribution

$$\pi(\lambda|\mathbf{N}) \propto \frac{(\lambda/\beta)^{\alpha-1}}{\Gamma(\alpha)\beta} e^{-\lambda/\beta} \prod_{i=1}^{n} e^{-\lambda} \frac{\lambda^{N_i}}{N_i!} \propto \lambda^{\hat{\alpha}-1} e^{-\lambda/\hat{\beta}} \tag{11}$$

is $Gamma(\hat{\alpha}, \hat{\beta})$ with updated parameters

$$\hat{\alpha} = \alpha + \sum_{i=1}^{n} N_i \text{ and } \hat{\beta} = \beta/(1+\beta \times n). \tag{12}$$

If the prior distribution is non-informative improper constant, then using (5), the posterior distribution is $Gamma(\hat{\alpha}, \hat{\beta})$, with parameters

$$\hat{\alpha} = \sum_{i=1}^{n} N_i + 1 \text{ and } \hat{\beta} = 1/n. \tag{13}$$

In this case, the mode of the posterior distribution $\pi(\lambda|\mathbf{N})$ is $\hat{\lambda} = (\hat{\alpha}-1)\hat{\beta} = \frac{1}{n}\sum_{i=1}^{n} N_i$ which is the same as MLE.

## Lognormal distribution (severity)

Suppose that, given $\mu$ and $\sigma$, the severities $\mathbf{X} = (X_1,...,X_n)$ are independent random variables from Lognormal distribution $LN(\mu,\sigma)$ with a density

$$f(x|\mu,\sigma) = \frac{1}{x\sqrt{2\pi\sigma^2}} \exp\left(-\frac{(\ln x - \mu)^2}{2\sigma^2}\right). \tag{14}$$

That is, $Y_i = \ln X_i$, $i = 1,...,n$ are independent and identically distributed from the Normal density $N(\mu,\sigma)$. Denote $\overline{Y} = \frac{1}{n}\sum_{i=1}^{n} Y_i$ and $\overline{Y^2} = \frac{1}{n}\sum_{i=1}^{n} Y_i^2$. Assume the following joint prior distribution $\pi(\mu,\sigma^2) = \pi(\mu|\sigma^2)\pi(\sigma^2)$, where

$$\pi(\sigma^2) = InvChiSq(\nu,\beta) = \frac{2^{-\nu/2}}{\beta\Gamma(\nu/2)} \left(\frac{\sigma^2}{\beta}\right)^{-\frac{\nu}{2}-1} \exp\left(-\frac{\beta}{2\sigma^2}\right),$$

$$\pi(\mu|\sigma^2) = N(\theta,\sigma^2/\phi) = \frac{1}{\sqrt{2\pi\sigma^2/\phi}} \exp\left(-\frac{(\mu-\theta)^2}{2\sigma^2/\phi}\right). \tag{15}$$

Here, $InvChiSq(\nu,\beta)$ is the Inverse Chi-squared density. Then the posterior is

$$\pi(\mu,\sigma^2|\mathbf{X}) = \pi(\mu|\sigma^2,\mathbf{X})\pi(\sigma^2|\mathbf{X}) \text{ with}$$

$$\pi(\mu|\sigma^2,\mathbf{X}) = N(\hat{\theta},\sigma^2/\hat{\phi}), \quad \pi(\sigma^2|\mathbf{X}) = InvChiSq(\hat{\nu},\hat{\beta}) \tag{16}$$

and parameters

$$\hat{v} = v + n, \quad \hat{\beta} = \beta + \phi\theta^2 + n\overline{Y^2} - (\phi\theta + n\overline{Y})^2/(\phi + n),$$
$$\hat{\theta} = (\phi\theta + n\overline{Y})/(\phi + n), \quad \hat{\phi} = \phi + n. \qquad (17)$$

The marginal posterior distribution for $\mu$ is shifted $t$-distribution with $\hat{v}$ degrees of freedom:

$$\pi(\mu | \mathbf{X}) = \int \pi(\mu, \sigma^2 | \mathbf{X}) d\sigma^2 \propto \left[1 + \frac{1}{\hat{v}}\left(\frac{\mu - \hat{\theta}}{\hat{\gamma}}\right)^2\right]^{-\frac{\hat{v}+1}{2}}, \qquad (18)$$

where $\hat{\gamma} = \sqrt{\hat{\beta}/(\hat{\phi}\hat{v})}$. Assuming non-informative constant priors, the posterior distributions are $\pi(\sigma^2 | \mathbf{X}) = InvChiSq(\hat{v}, \hat{\beta})$ and $\pi(\mu | \sigma^2, \mathbf{X}) = N(\hat{\theta}, \sigma^2/\hat{\phi})$ with parameters

$$\hat{v} = n - 3, \quad \hat{\beta} = n\overline{Y^2} - n(\overline{Y})^2, \quad \hat{\theta} = \overline{Y} \text{ and } \hat{\phi} = n. \qquad (19)$$

In this case, the mode of $\pi(\mu, \sigma^2 | \mathbf{X})$ is $\hat{\mu} = \overline{Y}$ and $\hat{\sigma}^2 = \overline{Y^2} - (\overline{Y})^2$ which are equivalent to the maximum likelihood estimators.

## Pareto distribution (severity)

Another important example of the severity distribution, often used to fit the tail of the severity distribution for a given threshold $L > 0$, is the Pareto distribution with a density

$$f(x|\xi) = \frac{\xi}{L}\left(\frac{x}{L}\right)^{-\xi-1}. \qquad (20)$$

It is defined for $x \geq L$ and $\xi > 0$. If $\xi > 1$, then the mean is $L\xi/(\xi - 1)$, otherwise the mean does not exist. Suppose that conditionally, given $\xi$, severities $\mathbf{X} = (X_1, ..., X_n)$ are independent random variables from the Pareto distribution and the prior for the tail parameter $\xi$ is $Gamma(\alpha, \beta)$. Then the posterior distribution is $Gamma(\hat{\alpha}, \hat{\beta})$ with

$$\hat{\alpha} = \alpha + n \text{ and } \hat{\beta}^{-1} = \beta^{-1} + \sum_{i=1}^{n}\ln(X_i/L). \qquad (21)$$

If the tail parameter $\xi$ has constant non-informative prior, then the posterior distribution is $Gamma(\hat{\alpha}, \hat{\beta})$ with parameters

$$\hat{\alpha} = n + 1 \text{ and } \hat{\beta}^{-1} = \sum_{i=1}^{n}\ln(X_i/L). \qquad (22)$$

In this case, the mode of posterior distribution is achieved at $\hat{\xi} = n/\sum_{i=1}^{n}\ln(X_i/L)$, which is the same as MLE.

It is interesting to note that assumption of $\xi > 0$ leads to the predicted annual loss distribution (2) whose mean is infinite because $\Pr[\xi \leq 1]$ is finite. If we do not want to allow for infinite mean predicted loss, then the prior distribution for parameter $\xi$ should be restricted to $\xi > 1$.

## Numerical Examples

Based on the framework described above, we perform numerical example to demonstrate the impact of parameter uncertainty as follows:

- Simulate losses over $M$ years from $Poisson(\lambda_0 = 10)$ and $LN(\mu_0 = 1, \sigma_0 = 2)$, i.e. $\mathbf{Y} = (N_1,...,N_M, X_1,...,X_K)$, $K = \sum_{m=1}^{M} N_m$.
- Given simulated losses, calculate point estimators for distribution parameters $\hat{\lambda}$, $\hat{\mu}$ and $\hat{\sigma}$ using the maximum likelihood method. Find the 0.999 quantile $\hat{Q}_{0.999}$ of the conditional annual loss distribution $g(Z | \hat{\lambda}, \hat{\mu}, \hat{\sigma})$, using the MC method with $10^6$ simulations.
- Find the posterior distributions for parameters $\pi(\lambda, \mu, \sigma | \mathbf{Y})$ assuming non-informative constant priors using formulas given in the above sections. Calculate the 0.999 quantile $\hat{Q}^B_{0.999}$ of the predictive distribution (2), using MC method with $10^6$ simulations. Here, we use non-informative constant priors to ensure that all inferences are based on the observed losses only (one can also use constant priors defined on a very wide interval).
- Estimate bias $E[\hat{Q}^B_{0.999} - \hat{Q}_{0.999}]$ by repeating the above steps 100 times.

The parameter uncertainty is ignored by the estimator $\hat{Q}_{0.999}$ but is taken into account by $\hat{Q}^B_{0.999}$. In Figure 1 and Table 1, we present results for $\hat{Q}_{0.999}$ and $\hat{Q}^B_{0.999}$ versus number of years $M$ for one of the simulated realizations. Figure 2 shows the relative bias (calculated over 100 realizations) $E[\hat{Q}^B_{0.999} - \hat{Q}_{0.999}]/Q^{(0)}$, where $Q^{(0)}$ is the quantile of the $g(Z | \boldsymbol{\theta}_0)$, when severities are simulated from $LN(\mu_0 = 1, \sigma_0 = 2)$ and $Pareto(\xi_0 = 2, L = 1)$. The estimator $\hat{Q}^B_{0.999}$ converges to $\hat{Q}_{0.999}$ from above as the number of losses increases. $\hat{Q}^B_{0.999}$ is significantly larger than $\hat{Q}_{0.999}$ for small number of losses. The bias induced by parameter uncertainty is large: it is approximately 10% after 40 years (i.e. approximately 400 data points) in our example (MC numerical standard errors for calculated quantiles are 1-2%).

The above numerical examples are given for illustrative purposes only. The parameter values used in the example may not be typical for many operational risks. One should do the above analysis with real data to find the impact of parameter uncertainty. For high frequency low impact risks, where a large number of data is available, the impact is certainly expected to be small. However for low frequency high impact risks, where the data are very limited, the impact can be significant.

# Conclusions

In this paper we have described the use of the Bayesian inference method for the quantification of the operational risk accounting for parameter uncertainty. Under this framework, the capital charge is estimated using the 0.999 quantile of the predictive distribution. This estimator is larger then the point estimator of the 0.999 quantile of the annual loss distribution calculated using the maximum likelihood parameter estimates, reflecting the fact that extra capital is required to cover for parameter uncertainty. The bias induced by the parameter uncertainty can be significant. Only for a large number of observations the bias can be neglected.

In our examples, non-informative constant prior distributions for the parameters are used to ensure that all inferences are based on the observed losses only (one can also use constant priors defined on a very wide interval). Specifying informative priors (using expert opinions or external data) may significantly reduce the bias due to parameter uncertainty and allows to combine internal data, expert opinions and external data simultaneously, see Lambrigger, Shevchenko and Wüthrich (2007).

MC estimation of the predictive distribution and its 0.999 quantile is computationally very similar (in terms of time and complexity) to the calculation of the annual loss distribution conditional on parameter point estimates (assuming that the posterior distribution is known as in our examples). If the posterior distribution is not known in closed form then available MCMC algorithms or approximation (7) can be used efficiently to estimate the posterior.

Another good feature of the proposed approach, appealing to risk managers, is that the 0.999 quantile of the full predictive annual loss distribution (2) provides a final point estimator for the capital charge accounting for parameter uncertainty rather than a confidence interval for the 0.999 quantile. The later case can be more computationally demanding (if the MC method is used) and requires to choose a confidence level for the interval.

# Acknowledgements

The author would like to thank Mario Wüthrich, Doug Shaw and Peter Thomson for fruitful discussions and useful comments.# References

Berger, J. O. (1985). *Statistical Decision Theory and Bayesian Analysis*. Springer-Verlag, New York.

BIS (2006). *Basel II: International Convergence of Capital Measurement and Capital Standards: a Revised Framework-Comprehensive Version*. Bank for International Settlements (BIS), www.bis.org.

Frachot, A., Moudoulaud, O. and Roncalli, T. (2004). Loss distribution approach in practice. *The Basel Handbook: A Guide for Financial Practitioners*, edited by M. Ong, Risk Books.11

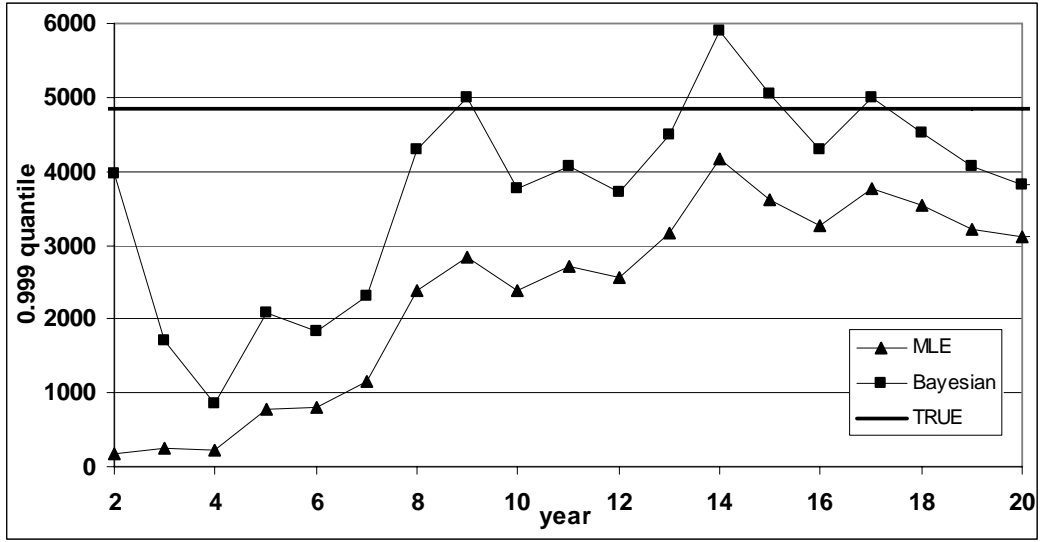

**Figure 1.** Comparison of the 0.999 quantile estimators. Parameter uncertainty is ignored by $\hat{Q}_{0.999}$ (MLE) but is taken into account by $\hat{Q}^{B}_{0.999}$ (Bayesian). Losses and frequencies were simulated from *Poisson*(10) and *LN*(1,2). Non-informative constant prior are assumed for calculation of the posterior distribution.

**Table 1.** Losses and frequencies were simulated from *Poisson*(10) and *LN*(1,2) over *M* years. *K* is the total number of losses simulated over *M* years. $\hat{\lambda}$, $\hat{\mu}$ and $\hat{\sigma}$ are the point MLEs. 0.95 predictive interval for the parameters, calculated using posterior distribution, is given in brackets. $\hat{Q}^{B}_{0.999}$ is the 0.999 quantile estimator accounting for parameter uncertainty. $\hat{Q}_{0.999}$ is the quantile estimator calculated using point MLEs. The 0.999 quantile for true parameters is $\approx 4.9$. All quantile estimates are in thousands.

| *M* | *K* | $\hat{\mu}\ (\mu_L,\mu_U)$ | $\hat{\sigma}\ (\sigma_L,\sigma_U)$ | $\hat{\lambda}\ (\lambda_L,\lambda_U)$ | $\hat{Q}_{0.999}$ | $\hat{Q}^{B}_{0.999}$ |
|---|---|---|---|---|---|---|
| 5 | 43 | 0.08 (-0.49,0.64) | 1.76(1.50,2.34) | 8.60(6.39,11.58) | 0.8 | 2.1 |
| 10 | 101 | 0.42 (0.03,0.82) | 1.97(1.75,2.32) | 10.10(8.32,12.27) | 2.4 | 3.8 |
| 15 | 150 | 0.62 (0.29, 0.95) | 2.03(1.83,2.31) | 10.00(8.53,11.73) | 3.6 | 5.0 |
| 20 | 204 | 0.68(0.40, 0.95) | 1.97(1.80,2.20) | 10.20(8.89,11.70) | 3.1 | 3.8 |
| 40 | 391 | 0.87(0.67,1.07) | 1.97(1.85,2.13) | 9.78(8.85,10.79) | 3.7 | 4.1 |
| 60 | 609 | 0.86(0.70,1.03) | 1.99(1.88,2.11) | 9.60(8.85,10.42) | 3.9 | 4.2 |
| 80 | 797 | 0.91(0.78,1.05) | 1.93(1.85,2.04) | 9.96(9.29,10.68) | 3.5 | 3.6 |
| 100 | 1015 | 0.95(0.83,1.07) | 1.95(1.87,2.04) | 10.15(9.54,10.79) | 3.9 | 4.0 |
| 200 | 2019 | 0.97(0.89,1.06) | 1.98(1.92,2.04) | 10.10(9.66,10.55) | 4.4 | 4.5 |
| 400 | 4003 | 0.98(0.92,1.05) | 2.01(1.97,2.05) | 10.01(9.70,10.32) | 4.9 | 4.9 |

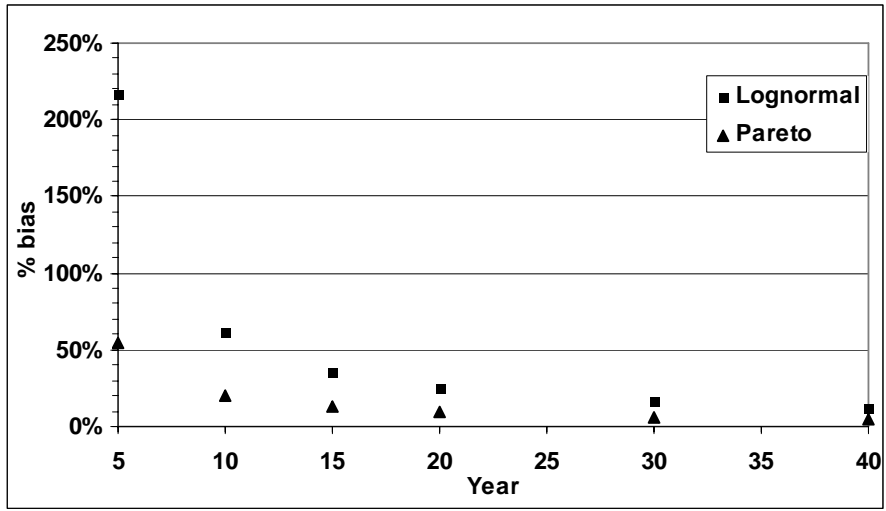

**Figure 2. Relative bias (calculated over 100 realizations) in the 0.999 quantile estimator induced by the parameter uncertainty vs the number of observation years. (Lognormal) - losses were simulated from *Poisson*(10) and *LN*(1,2). (Pareto) – losses were simulated from *Poisson*(10) and *Pareto*(2) with *L*=1.**